\documentclass[twocolumn,english,prb,showpacs,
superscriptaddress,preprintnumbers,amsmath,amssymb]{revtex4}
\usepackage{graphicx}
\begin{document}

\title{Thermodynamic properties of ferromagnetic mixed-spin chain systems }

\author{Noboru Fukushima}
\email{n.fukushima@tu-bs.de}
\author{Andreas Honecker}
\affiliation{Institut f\"{u}r Theoretische Physik, Technische Universit\"{a}t
Braunschweig, D-38106 Braunschweig, Germany }
\author{Stefan Wessel}
\affiliation{Theoretische Physik, ETH Z\"urich, CH-8093 Z\"urich, Switzerland}
\author{Wolfram Brenig}
\affiliation{Institut f\"{u}r Theoretische Physik, Technische Universit\"{a}t
Braunschweig, D-38106 Braunschweig, Germany }

\begin{abstract}
Using a combination of high-temperature series expansion, exact
diagonalization and quantum Monte Carlo, we perform a complementary
analysis of the thermodynamic properties of quasi-one-dimensional
mixed-spin systems with alternating magnetic moments.  In addition to
explicit series expansions for small spin quantum numbers, we present an
expansion that allows a direct evaluation of the series coefficients as
a function of spin quantum numbers. Due to the presence of excitations
of both acoustic and optical nature, the specific heat of a mixed-spin
chain displays a double-peak-like structure, which is more pronounced
for ferromagnetic than for antiferromagnetic intra-chain exchange.  We
link these results to an analytically solvable half-classical limit.
Finally, we extend our series expansion to incorporate the single-ion
anisotropies relevant for the molecular mixed-spin ferromagnetic chain
material MnNi(NO$_{2}$)$_{4}$(ethylenediamine)$_{2}$, with alternating
spins of magnitude 5/2 and 1. Including a weak inter-chain coupling, we
show that the observed susceptibility allows for an excellent fit, and
the extraction of microscopic exchange parameters.
\end{abstract}


\maketitle

\section{Introduction}

Promoted by the synthesis of various one-dimensional (1D) bimetallic
molecular magnets, the physics of quantum spin chains with mixed magnetic
moments is of great interest. Typically, quasi-1D mixed-spin (MS)
compounds display \textit{antiferromagnetic} (AFM) intra-chain exchange
\cite{Kahn95,Caneschi89,Nishizawa00}, which has stimulated theoretical
investigations of AFM MS models using a variety of techniques such
as spin-wave theory \cite{Pati97a,Pati97b,Brehmer97,Yamamoto98aa,Yamamoto98a,Yamamoto98b,Yamamoto98c},
variational methods \cite{Kolezhuk97}, the density-matrix renormalization
group (DMRG)\cite{Pati97a,Pati97b,Yamamoto98b}, and quantum Monte Carlo
calculations\cite{Brehmer97,Yamamoto98aa, Yamamoto98a,Yamamoto98b,Yamamoto98c}. Interestingly, since
a unit cell of the MS chain comprises of two different magnetic moments,
the spectrum will allow for excitation of `acoustic' as well as
`optical' nature \cite{Yamamoto98aa,Yamamoto98b}.
While not identified unambiguously in present day experiments, the
character of these excitations should appear in thermodynamic and
other observable properties as two \emph{independent energy scales}.
\cite{Nakanishi02}

In addition, and apart from the preceding, what remains less well
studied are MS chains with \emph{ferro}\textit{magnetic} (FM) intra-chain
exchange, which arise in materials of recent interest such as MnNi(NO$_{2}$)$_{4}$(en)$_{2}$
with en = ethylenediamine \cite{Feyerherm01}. This compound is regarded
as a quasi-1D MS material with spins $S=5/2$ and $s=1$ at the Mn
and Ni ions, respectively. The susceptibility displays an easy-axis
anisotropy. At temperatures below $T_{N}=2.45\textrm{K}$, a weak
AFM inter-chain coupling induces AFM ordering. If this antiferromagnetic
order is suppressed by a magnetic field of approximately 1.6T, the
low-temperature specific heat shows a maximum at $T=6\textrm{K}$
and a shoulder at $T=1.5\textrm{K}$. These features could possibly
reflect the two aforementioned characteristic energy scales, however
for the case of an FM, rather than an AFM MS chain.

Motivated by this, it is the purpose of this paper to perform a complementary
analysis of the thermodynamic properties of FM MS chains,
using high-temperature series expansion (HTSE),
exact diagonalization (ED), and quantum Monte Carlo (QMC).
In particular, our HTSE will be derived
for \textit{arbitrary} alternating spins $S$ and $s$. This will
allow not only for a direct comparison with experimental data, but
also for a study of a gradual transit to the `half-classical'
limit $S\rightarrow \infty $ which is exactly solvable \cite{Dembiski75,Seiden83,Kahn92}.
Finally, while our ED and QMC data will be obtained on systems of
the smallest possible mixed-spin magnitude, i.e. $S=1$ and $s=1/2$,
this is a
case also included in the HTSE. In addition, for the sake of completeness
and to compare with existing literature, we will also include results
for the AFM case.

In this paper, we focus on a chain with two kinds of spin species,
i.e. $S$ and $s$, arranged alternatingly and coupled by a nearest-neighbor
Heisenberg exchange. Namely, the Hamiltonian reads \begin{equation}
\mathcal{H}_{\textrm{int}}=-J\sum _{i=1}^{N}\left(\mathbf{S}_{i}\cdot \mathbf{s}_{i}+\mathbf{s}_{i}\cdot \mathbf{S}_{i+1}\right).\label{Hint}\end{equation}
The subscript $i=1\ldots N$ refers to the unit cells, and we always use
periodic boundary conditions.

In section \ref{sec:HTSE}, our HTSE approach is detailed. In section
\ref{sec:ClassEd}, we turn to a comparison with QMC and ED and the results
of the `half-classical' limit. In section \ref{sec:Fit}, we discuss
the result of fitting the HTSE to susceptibility data obtained for
MnNi(NO$_{2}$)$_{4}$(en)$_{2}$. Conclusions are presented in section
\ref{sec:Summary}.

\section{High-temperature series expansion\label{sec:HTSE}}

The HTSE is an expansion in powers of $\beta J$, where $\beta $ is the
inverse temperature. Here we use the linked-cluster expansion of Ref.~\onlinecite{Domb3}.
In this method, the series coefficients for the thermodynamic limit
are obtained \emph{exactly} from those calculated on finite-size clusters.
In general, this includes the subtraction of contributions from a
large number of so-called subclusters. In one dimension, however,
significant simplifications occur due to cancellation
\cite{Baker64,Fukushima03}. This is true also for the MS chain systems.
That is, in the absence of a magnetic field,
the free energy $F$ in the thermodynamic limit is represented by
\begin{eqnarray}
F/N & = & F_{\ell }(S,s)+F_{\ell }(s,S)\nonumber \\
 &  & -F_{\ell -1}(S,s)-F_{\ell -1}(s,S)+O[(\beta J)^{2\ell }],
\end{eqnarray}
 where $F_{\ell }(S,s)$ is the free energy of the $\ell $-site open-chain
system described by \begin{eqnarray}
\mathcal{H}_{\ell } & = & -J\sum _{i=1}^{(\ell -1)/2}\left(\mathbf{S}_{i}\cdot \mathbf{s}_{i}+\mathbf{s}_{i}\cdot \mathbf{S}_{i+1}\right)\; (\ell :\textrm{odd})\nonumber \\
 & = & \mathcal{H}_{\ell -1}-J\mathbf{S}_{\frac{\ell }{2}}\cdot \mathbf{s}_{\frac{\ell }{2}}\; \; (\ell :\textrm{even}),
\end{eqnarray}
 and $F_{\ell }(s,S)$ is obtained by exchanging $S$ and $s$ in
$F_{\ell }(S,s)$. Then, a calculation of $\textrm{Tr}[(\mathcal{H}_{\ell })^{n}]$
is needed on a finite system, i.e.,\begin{equation}
\textrm{Tr}\mathcal{H}_{\ell }^{n}=\sum _{\{m_{i}\}}\langle m_{1}\ldots m_{\ell }|\mathcal{H}_{\ell }^{n}|m_{1}\ldots m_{\ell }\rangle ,\label{eq:w1}\end{equation}
 where $m_{i}$ represents the magnetic quantum numbers at site $i$.
We apply $\mathcal{H}_{\ell }$ order by order on the ket
$|m_{1}\ldots m_{\ell }\rangle $. This operation yields linear combinations
of kets with coefficients which are functions of $\{m_{i}\}$.
To evaluate $\textrm{Tr}\mathcal{H}_{\ell }^{2n}$, products of kets of
type $\mathcal{H}_{\ell }^{n}|m_{1}\ldots m_{\ell }\rangle $ are needed
at most, while in the case of Tr$\mathcal{H}_{\ell }^{2n+1}$, one can
use that
$\mathcal{H}_{\ell }^{2n+1}=\mathcal{H}_{\ell}^{n}\mathcal{H}_{\ell }^{n+1}$.
In order to evaluate this trace, we use two different algorithms.

Method (i) is based on a direct matrix multiplication
for fixed $S$ and $s$.
A linear combination of kets with coefficients is regarded as
a sparse vector. It is stored as a compressed array
of non-zero elements and another array of their pointers to the kets.
These pointers are stored in the ascending order
so that one can find a needed element
using binary search in the array.
All the operations are performed using integers,
and thus there is no loss of precision.

Method (ii) is designed for arbitrary spins,
which is based on an analytic approach to the matrix
elements in Eq.~(\ref{eq:w1}).
It has an advantage for large spins because
the method (i) will fail for
very large spins due to time and/or memory constraints.
After symbolic operations,
the summation of the form $\sum _{m=-s}^{s}m^{n}$
can be calculated analytically for arbitrary $n$.
For example, when $n=2$, the sum is equal to $\frac{1}{3}s\, (s+1)(2s+1)$.
As a result, the series coefficients are obtained as an expression
valid for arbitrary $S$ and $s$.
Naively it may seem that operators of type $s^{\pm }$ will causes
square roots in the matrix elements of the Hamiltonian. However, such
square roots are absent in the final result. In fact, the calculation
can be carried out disregarding square roots as explained below. Introducing
the simplified notation, \begin{equation}
|\pm n_{1}\ldots \pm n_{\ell })\equiv (s_{1}^{\pm })^{n_{1}}\ldots (s_{\ell }^{\pm })^{n_{\ell }}|m_{1}\ldots m_{\ell }\rangle .\label{eq:newket}\end{equation}
 the initial state is represented by $|0\ldots 0)$. Spin operators
act on these states as follows. Suppose $|\pm n)$ ($n$>0) represents
the state at site $i$ in the notation (\ref{eq:newket}). Then,\begin{alignat}{1}
s^{z}|\pm n) & =(m\pm n)|\pm n),\\
s^{\pm }|\pm n) & =|\pm n\pm 1),
\end{alignat}
 and\begin{eqnarray}
s^{\mp }|\pm n) & = & s^{\mp }s^{\pm }|\pm n\mp 1)\nonumber \\
 & = & |\pm n\mp 1)\times \nonumber \\
 &  & \; \left\{ s(s+1)-(m+n\mp 1)(m+n)\right\} .
\end{eqnarray}
 Note that the norm of $|n)$ is not unity, namely, \begin{equation}
(\pm n|\pm n)=\prod _{n'=1}^{n}\left\{ s(s+1)-(m+n'\mp 1)(m+n')\right\} .\end{equation}

Besides the methods (i) and (ii), the contribution
to the specific heat from the
largest cluster is calculated separately.
Namely, contributions from
$\ell$-site chain to $O[(\beta J)^{2\ell-2}]$
and $O[(\beta J)^{2\ell-1}]$
have a simple form;
with notation $x\equiv s(s+1)$ and $X\equiv S(S+1)$,
for $\ell=2 l$ it is proportional to 2$l$ $x^l X^l$
and for $\ell=2 l+1$ to
$x^l X^{l+1}+x^{l+1} X^{l}$.
The prefactors of these terms
can be determined by comparing with those of $s=S=1/2$
for any $\ell$ in Ref.~\onlinecite{Fukushima03}.
The methods (i) and (ii) are used only for the rest
of the contribution.

We have computed the specific heat for the model (\ref{Hint})
with $s=1/2$ and $S=1$, up to 29th order using the method (i).
Furthermore, for arbitrary $s$ and $S$, the series
has been calculated
up to 11th order using the method (ii)
and standard symbolic packages \cite{Mathematica}.
By setting $S=s=1/2$, our model is reduced to the homogeneous spin-half
Heisenberg chain, and our series agrees with that in the literature\cite{Baker64}.
Furthermore, we have also checked that the $S\rightarrow \infty $
limit of the series agrees with the Taylor series of the exact solution\cite{Dembiski75}
which will be shown in the next section.

\section{large- and small-spin limits\label{sec:ClassEd}}

\begin{figure}
\begin{center}\includegraphics[  width=7.5cm,
  keepaspectratio]{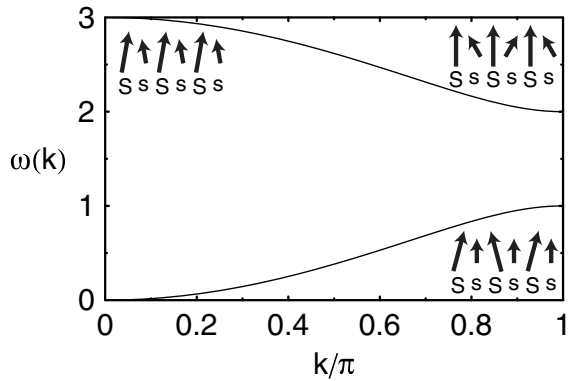}\end{center}

\caption{\label{fig:dispersion} The spin-wave dispersion relation
for the FM MS chain with $s=1/2$ and $S=1$. The spatial distance
of two spins of the same species in neighboring unit cells is taken
to be unity. }
\end{figure}

In this section, we provide evidence for the presence of a
double-peak-like structure in the specific heat of the FM MS chain.
We begin by recalling the elementary excitations in an MS chain. The dispersion relation of the one-magnon
excitations in the FM case reads \begin{equation}
\omega (k)=J\left(S+s\pm \sqrt{S^{2}+s^{2}+2Ss\cos (k)}\right),\end{equation}
and is shown in Fig.~\ref{fig:dispersion}
for the extreme quantum case, $s=1/2$ and $S=1$. Similar to the AFM case\cite{Yamamoto98aa,Yamamoto98b}, the
spectrum consists of both an acoustic and an
optical branch reflecting the presence of two different spins in a unit cell.
These branches
indicate two energy scales in the thermodynamics of the MS chains.
The appearance of two one-magnon branches can be
qualitatively understood as follows:
At $k=\pi$, the excitations correspond
to an alternating tilting of either the shorter or the longer spins
as indicated by the arrows in Fig.~\ref{fig:dispersion}.
The magnitude of neighboring spins determines the energies of these modes,
leading to a splitting of the branches for
$S \ne s$. As $k$ is reduced away from $k=\pi$,
the magnon eigenstates gradually loose these alternating tilting form,
and near $k=0$ become similar in nature to those of uniform chains.
The low-temperature specific heat as computed using linear spin-wave
theory is expected to be exact\cite{Takahashi85} to leading order in $T$. Since $\omega (k)$
is proportional to $k^{2}$ at low energies, the specific heat is thus
proportional to $T^{1/2}$ for $T\ll J$.

We can establish a
connection between the high-temperature series for the specific heat and this low-temperature scaling using
a suitable Pad\'{e} analysis.
In extrapolating the high-temperature
series for the specific heat, information from the ground-state energy,
the low-energy excitations, and the high-temperature entropy can be used
as follows \cite{Bernu01}: The expansion variable is changed from
$\beta $ to the internal energy per unit cell, $e=\langle \mathcal{H}_{\textrm{int}}\rangle /N$.
Here, $e=0$ for $\beta =0$. The power series of $e(\beta )$ is thus
inverted to obtain $\beta (e)$. Let $\mathcal{S}(e)$ denote the entropy per unit cell of the MS chain.
{}From $\beta =\textrm{d}\mathcal{S}/\textrm{d}e$, one obtains
\begin{equation}
\mathcal{S}=\mathcal{S}_{T=\infty }+\int _{0}^{e}\beta
(e')\textrm{d}e',\end{equation} where $\mathcal{S}_{T=\infty }=\ln
(2S+1)+\ln (2s+1)$. The low-temperature behavior of the specific
heat, $C\propto T^{1/2}$, translates into $\mathcal{S}\propto
(e-e_{0})^{1/3}$ at $e\sim e_{0}$ for the new series, where $e_{0}$ is
the ground state energy per site.  In
Ref.~\onlinecite{Bernu01}, only the FM $e_0$ is used to extrapolate the
specific heat of a HTSE for a FM model. Here, we also employ the AFM $e_0$, since this additional constraint
from the other sign of $J$ does not drastically change the final
result, but makes the extrapolation less sensitive to the used Pad\'{e} approximant.
We thus obtain the same extrapolation  for both the FM and the AFM case.
Namely, the AFM specific heat will be obtained using the substitutions $e\rightarrow -
e$ and $\beta J \rightarrow - \beta J$ from the FM case.  Then, if the
Pad\'{e} approximant in $e$ is applied to \[ \mathcal{S}^{3}\Big (\ln
(2S+1)(2s+1)\Big ) ^{-3}\left(1-\frac{e}{e_{0}^{\rm FM}}\right)^{-1}
\left(1+\frac{e}{e_{0}^{\rm AFM}}\right)^{-1},\] the low-temperature
behavior and the high-temperature entropy are correctly reproduced.
For $s=1/2$ and $S=1$, we use the AFM ground-state energy  $e_{0}^{\rm
AFM}=-1.45408J$ from Ref.~\onlinecite{Pati97a}. The extrapolation
is found to be rather insensitive to errors in $e_{0}^{\rm AFM}$.
For example, an error $10^{-4}J$ in $e_{0}^{\rm AFM}$ affects the AFM specific heat
only by $4\times 10^{-4}$ at $T=0.1J$, and even less at higher temperatures.
In the following we denote by $P[m/n]$ the rational approximant
function in $e$ resulting from a polynomial of order $m$ over a
polynomial of order $n$.

\begin{figure}
\begin{center}\includegraphics[  width=8.5cm,
  keepaspectratio]{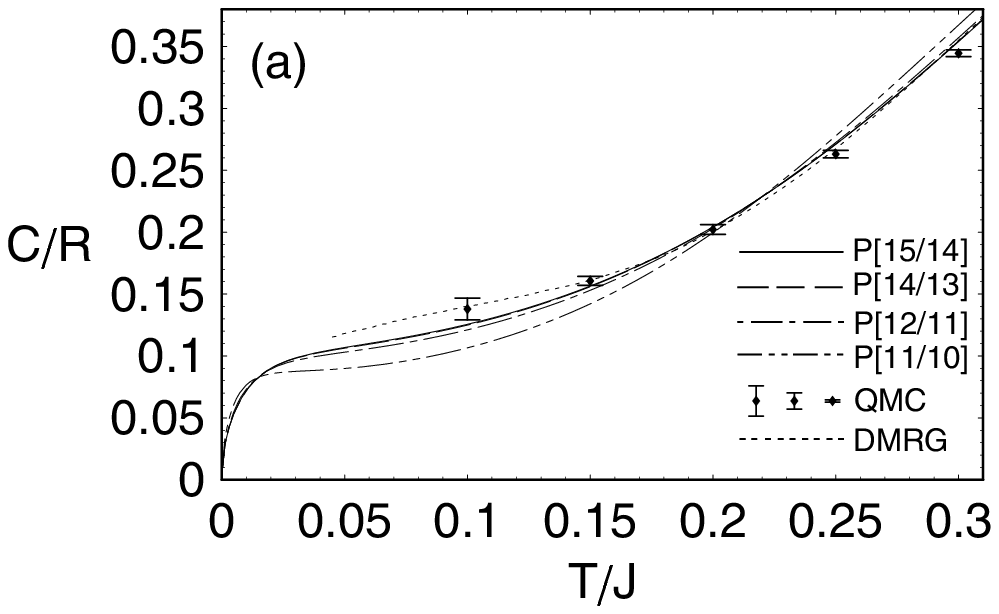}\end{center}
\begin{center}\includegraphics[  width=8.5cm,
  keepaspectratio]{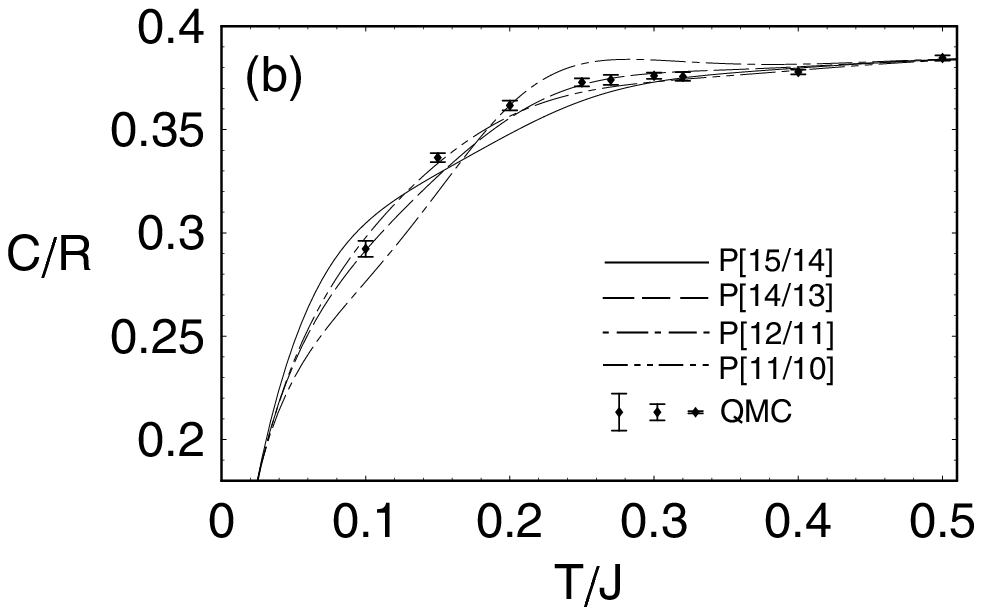}\end{center}

\caption{\label{fig:pade}
Low-temperature behavior of the specific heat per unit cell
of the AFM (a) and FM (b) mixed-spin chain with  $s=1/2$, $S=1$ using
several different Pad\'{e} approximants of the HTSE and from QMC.
In (a), DMRG results from Ref.~\onlinecite{Yamamoto98b}
are also shown. Here, $R$ is the gas constant. }
\end{figure}

In Fig.~\ref{fig:pade}, a comparison is shown between different Pad\'{e}
approximants for the low-temperature specific heat
for the $s=1/2$, $S=1$ case.  We also include results from
stochastic series expansion QMC simulations \cite{wessel}
for both the FM and the AFM case for chains of 100 sites.
No deviations were found to the QMC data for 50 sites, and thus we regard these
results to represent the thermodynamic limit.
For the AFM case, the DMRG results from Ref.~\onlinecite{Yamamoto98b} are shown in Fig.~\ref{fig:pade}a as well.
In contrast to the AFM case, the HTSE result for the low-temperature specific heat for the FM case
shows large oscillation upon increasing  the order of the series.
Comparison with the QMC data shows that the exact solution is located within the range of these
oscillations. We take the arithmetic average of $P[12/11]$, $P[13/12]$, $P[14/13]$ and
$P[15/14]$ as the final HTSE result, which in the
temperature range $0.03J<T<0.45J$ has an error of order 5\%.

\begin{figure}
\begin{center}\includegraphics[  width=8.5cm,
  keepaspectratio]{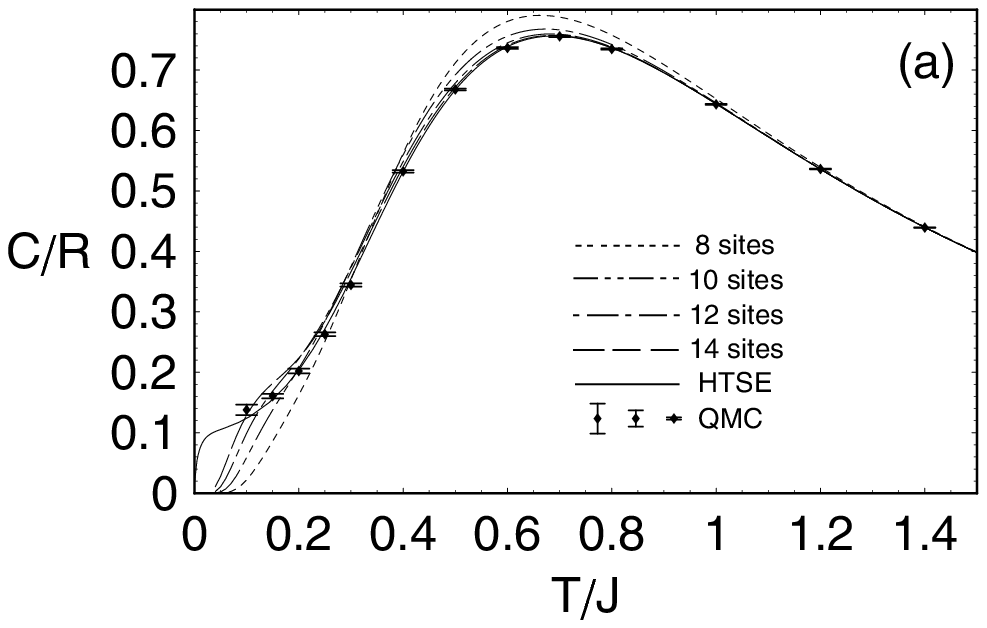}\end{center}

\begin{center}\includegraphics[  width=8.5cm]{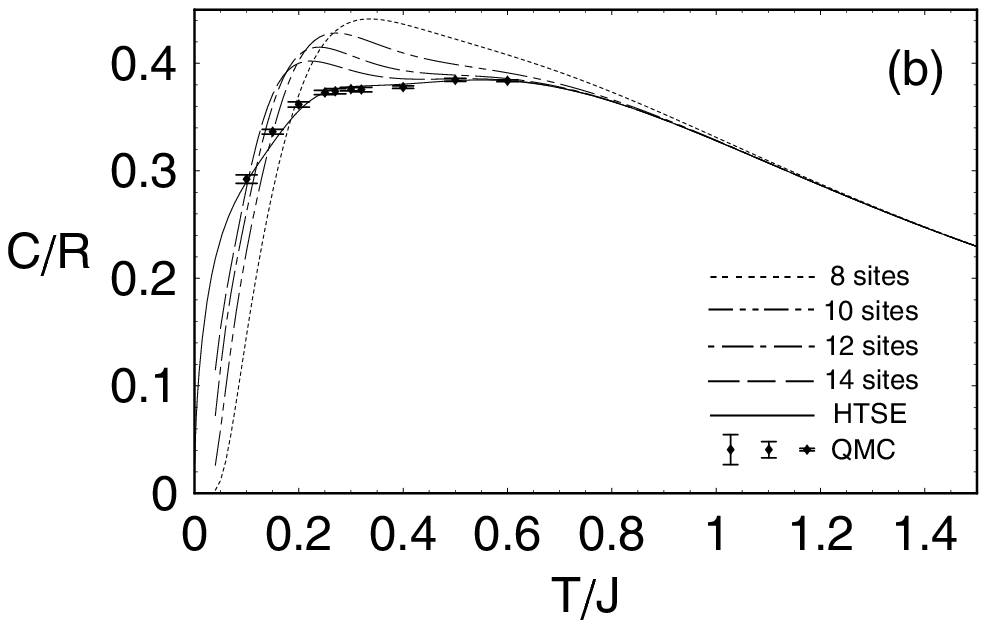}\end{center}

\caption{\label{fig:edhine}
Specific heat per unit cell
of the AFM (a) and FM (b) mixed-spin chain with  $s=1/2$, $S=1$ as obtained from
HTSE, QMC and full diagonalizations for small finite chains.
}
\end{figure}

The specific heat in a larger temperature regime is shown in Fig.~\ref{fig:edhine}.
We find overall good agreement between the QMC and the HTSE data,
both for the AFM and the FM case.
In contrast to the homogeneous FM spin-1/2 chain, the FM MS chain
displays at least two distinct structures in $C(T)$, namely a peak at $T\sim 0.54J$ and a shoulder at $T\sim 0.25J$.
The HTSE result indicates the presence of an additional weak shoulder at $T\sim 0.1J$, which is however
difficult to check for using QMC due to increasing statistical errors at that low temperatures.

The features in the intermediate temperature regime become more pronounced with increasing
order of the series expansion, probably because
the higher-order polynomials can reproduce the involved rapid changes more accurately.

We have included in Fig.~\ref{fig:edhine} ED results for small finite chains
for both the AFM and the FM case.
The full spectrum has been calculated for finite chains with
up to 7 unit-cells (14 sites).
The ED result for 14 sites agrees with the QMC and the HTSE
down to temperatures of $T\sim 0.5J$ for
the FM case, and $T\sim 0.3 J$ for the AFM case.
Fig.~\ref{fig:edhine} clearly shows that finite-size effects in the specific heat
are significantly larger in the FM than in the AFM case.
In the AFM case, the differences between the
12- and 14-site data are already rather small, with a weak shoulder
at low temperature signaling the second energy scale in $C\left(T\right)$.
For the FM case the ED data show a strong finite-size shift of the specific heat
maximum, which gradually develops into the shoulder near $T\sim 0.25J$,
which is found both from the HTSE and QMC.

This suggests that long-range collective modes
dominate the thermodynamics up to higher temperatures for the FM
than for the AFM case. To shed more light onto this difference,
let us compare the elementary excitation spectra for the FM and the AFM case:
(i) At low temperatures, the specific heat reflects the dispersion of the
acoustic branch of excitations. In the long-wavelength limit, the dispersion
in linear spin-wave theory reads\[
\omega (k)\sim \frac{Ss}{2|S\pm s|}k^{2},\]
 where the plus (minus) sign represents the FM (AFM\cite{Pati97a,Brehmer97}) case.
Therefore, the slope of the acoustic branch in the AFM case is larger
than in the FM case, making low-temperature shoulders more pronounced for
FM MS chains.
(ii) The gap between the acoustic and the optical
mode in the FM case is larger than in the AFM\cite{Yamamoto98aa,Yamamoto98b}
case, so that the splitting of the structures in $C(T)$ is
more evident in the former case. Apart  from (i) and (ii), the elementary
excitation spectra of the FM and the AFM case resemble each other,
pointing towards additional effects from magnon interactions.

After analyzing the extreme quantum case with $s=1/2$ and $S=1$, we now discuss
the specific heat of MS chains for larger values of $S$, keeping $s=1/2$ fixed. In particular, we
want to connect to the exactly solvable \cite{Dembiski75} `half-classical' limit, $S\rightarrow \infty $.

\begin{figure}
\begin{center}\includegraphics[  width=8.5cm,
  keepaspectratio]{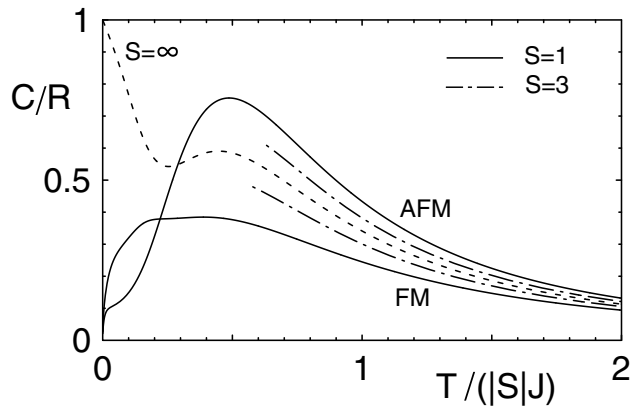}\end{center}

\caption{\label{fig:sinf} The HTSE results for the dependence on $S$ of
the specific heat per unit cell with $s=1/2$ for antiferromagnetic (AFM)
and ferromagnetic (FM) coupling, where
$|\mathbf{S}|=\sqrt{S(S+1)}$. Here, $S=\infty $ is from
Ref.~\onlinecite{Dembiski75}.  Details are in the text.}
\end{figure}

Fig.~\ref{fig:sinf} shows
the results from the 11th order HTSE for $s=1/2$ and various values of $S$.
The temperature is normalized to $|\mathbf{S}|J\equiv J\sqrt{S(S+1)}$
in order to render the 'half-classical' limit finite. In the limit  $S\rightarrow \infty $
the specific heat of the chain with FM couplings is identical to that
of the AFM chain, with a distinct peak at $T\sim 0.5\, |\mathbf{S}| J$.
At $T=0$ the specific heat of the `half-classical' model is finite.
In the quantum case, however, the specific heat vanishes as $T\rightarrow 0$.
Using the variable $S$ HTSE, we can link both limiting situations as follows:
In the limit $S\rightarrow \infty $, the series coefficients of $(\beta SJ)^{n}$
with odd $n$ converge to zero. Let us compare the $(2n)$-th and
$(2n+1)$-th order terms at finite $S$. The ratio of the latter to
the former is of $O(t/S)$, where $t\equiv T/(SJ)$. Hence, if the
limit $S\rightarrow \infty $ is taken with $t$ fixed, the latter
becomes negligible compared to the former as $S\rightarrow \infty $.
However, if $S$ is finite, the two terms are comparable for temperatures
below $t\sim S^{-1}$. Therefore, even if $S$ is  large,
the specific heat deviates from the 'half-classical' behavior
below $t\sim S^{-1}$, and approaches zero as $t\rightarrow 0$. As
a result, the specific heat will display a double-peak-like structure
for large $S$ as well.
Summarizing these results from the HTSE, there are two different ways of taking the
large-$S$ and  $T\rightarrow 0$ limit, \begin{eqnarray}
\lim _{T\rightarrow 0}\lim _{S\rightarrow \infty }C/R & = & 1,\\
\lim _{S\rightarrow \infty }\lim _{T\rightarrow 0}C/R & = & 0.
\end{eqnarray}

As seen from Fig.~\ref{fig:sinf} the difference between the FM and
the AFM cases are most pronounced in the extreme quantum limit
$s=1/2$ and $S=1$, and in the low-temperature regime.
The large- and small-spin limits exhibit similar structures,
suggesting that the specific heat of MS chain systems in general show
a double-peak-like or peak-shoulder structure, both
for AFM and FM intra-chain exchange, and for any combination of spins.
In the FM case this structure is more pronounced if $|S-s|$ is large,
reflecting the size of the gap between the acoustic and the optical mode,
similar to the AFM case.\cite{Nakanishi02}

We note that the specific heat in the high-temperature limit of the
AFM case is larger than in the FM case because of quantum effects.
This is evident from the first two terms of the series for the specific
heat, which we present in Appendix \ref{sec:seriesdata}. These two
terms stem from two-site correlations only and dominate the high-temperature
behavior.
Since the total entropy difference between zero
and infinite temperature is the same in both cases,
the FM and AFM specific-heat curves therefore have to intersect
at low temperatures.

\section{fitting to experimental data\label{sec:Fit}}

We now turn to a comparison to the susceptibility data observed on
MnNi(NO$_{2}$)$_{4}$(en)$_{2}$. In this compound, the symmetry
around Ni ions is nearly cubic with however a fairly large anisotropy
at the Mn site to be expected \cite{Feyerherm01}. Hence, we take
into account a single-site anisotropy only on one of the spins, i.e.
$S$. The \textit{g}-factors of the spins $S$ and $s$ are represented
by $G$ and $g$, respectively. Therefore, the total Hamiltonian reads
\begin{equation}
\mathcal{H}=\mathcal{H}_{\textrm{int}}+\mathcal{H}_{\textrm{ani}}+\mathcal{H}_{\textrm{mag}},\end{equation}
 where \begin{eqnarray}
\mathcal{H}_{\textrm{ani}} & = & D\sum _{i=1}^{N}\left(S_{i}^{z}\right)^{2},\\
\mathcal{H}_{\textrm{mag}} & = & -\mathbf{h}\cdot \sum _{i=1}^{N}(G\, \mathbf{S}_{i}+g\, \mathbf{s}_{i}).
\end{eqnarray}
 Here, we derive the power series of $\chi $ in $\beta D$ as well
as $\beta J$ up to $O(\beta ^{7})$. When $D=0$ and $g=G$,
the series coefficients
coincide with those in the literature\cite{Wojtowicz67,Freeman72}
assuming a misprint\cite{misprint}
in Ref.~\onlinecite{Wojtowicz67}.
Since the contribution $\mathcal{H}_{\textrm{mag}}$ is used to evaluate
the susceptibility we will only consider the case of small Zeeman
energies $\left|\mathbf{h}\right|=h\rightarrow 0$ in the following.
The orientation of the magnetic field $\mathbf{h}$ will be chosen
to be either $\mathbf{h}\Vert z$ or $\mathbf{h}\Vert x$. Most theoretical
studies of MS systems are limited to the case of $D=0$ and $g=G$,
in order to make use of total spin-$z$, i.e. $\sum _{i}(S_{i}^{z}+s_{i}^{z})$,
conservation. However, for a proper comparison to experimental data,
$D\neq 0$ and $g\neq G$ has to be accepted, leading to \begin{equation}
\left[\mathcal{H}_{\textrm{int}}+\mathcal{H}_{\textrm{ani}},\mathcal{H}_{\textrm{mag}}\right]\neq 0.\label{notcommute}\end{equation}
 We emphasize, that our HTSE is carried out taking into account this
non-commutativity.

\begin{figure}
\begin{center}\includegraphics[  width=8.5cm]{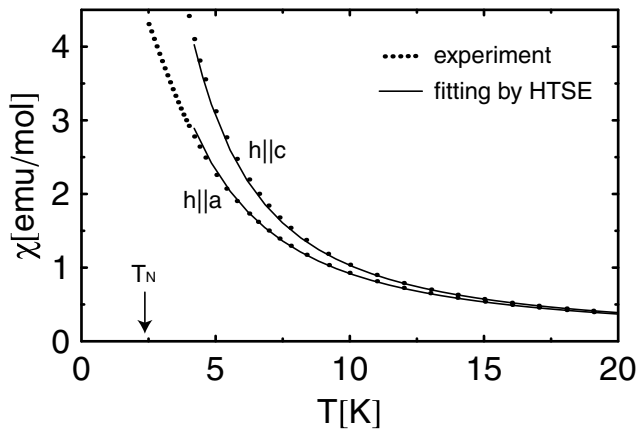}\end{center}

\caption{\label{fig:fitting}Fitting to susceptibility of MnNi(NO$_{2}$)$_{4}$(en)$_{2}$
by Eq.(\ref{Q1Dsus}). Here, $(\mu _{\textrm{B}}^{2}N_{\textrm{A}}/k_{\textrm{B}})\chi$
is plotted, where $\mu _{\textrm{B}}$, $N_{\textrm{A}}$ and $k_{\textrm{B}}$
are Bohr magneton, Avogadro's number and Boltzmann constant, respectively.}
\end{figure}

Since 3D AFM ordering of MnNi(NO$_{2}$)$_{4}$(en)$_{2}$ below $T_{\textrm{N}}=2.45\textrm{K}$
signals the presence of a non-negligible inter-chain exchange, we
will enhance our 1D analysis to incorporate this coupling on a phenomenological
basis. That is to say, fits to the experimental results will be performed
using an RPA expression \begin{equation}
\chi \simeq \frac{\chi _{\textrm{1D}}}{1-J_{\perp }\chi _{\textrm{1D}}},\label{Q1Dsus}\end{equation}
 where $\chi _{\textrm{1D}}$ is the susceptibility of the pure 1D
system obtained by HTSE and extrapolated by a simple Pad\'{e} approximation
(PA). Here, $J_{\perp }$ effectively models the average inter-chain
exchange. Figure \ref{fig:fitting} shows the results of our fits
of $\chi $ to experimental data of the susceptibility\cite{Feyerhermpri}
with a magnetic field oriented both, perpendicular and parallel to
the $c$-axis. Apart from $s=1$ and $S=5/2$ we have used $g=2.24$
and $G=2$ as listed in Ref.~\onlinecite{Feyerherm01}. Best fits
are obtained for $J=2.8\, \textrm{K}$, $J_{\perp }=-0.036\, \textrm{K}$
and $D=-0.36\, \textrm{K}$. As estimated from the PA of the HTSE,
in the range plotted, the error involved in the theoretical curve
for $\chi _{\textrm{1D}}$ is within the width of the line. As is
obviously from the figure our theory allows for an excellent fit to
the experimental data down to $T\sim $4K. Only the high-temperature
data, for $10\textrm{K}\leq T\leq 25\textrm{K}$ have been utilized
to set the parameters $J$, $D$, and $J_{\perp }$. Keeping these
fixed, the splitting between $\mathbf{h}\Vert c$ and $\mathbf{h}\perp c$
of the experimental data tends to become larger than that of the theory
for $T<4\textrm{K}$.

In Ref.~\onlinecite{Feyerherm01}, values of $J\sim 1.9\textrm{K}$
and $D\sim - 0.45\textrm{K}$ have been reported for MnNi(NO$_{2}$)$_{4}$(en)$_{2}$.
These have been obtained by fitting a directional average of the susceptibility
with respect to the magnetic field to the `half-classical' limit,
\cite{Kahn92} i.e. $S\rightarrow \infty $, applicable for $D=0$.
Moreover the effects of interchain coupling have been neglected. To
compare with this result we may use $J_{\perp }\equiv 0$ in Eq.~(\ref{Q1Dsus}).
In fair agreement with Ref.~\onlinecite{Feyerherm01}, we find $J\sim 2.4\textrm{K}$
and $D\sim - 0.36\textrm{K}$ in this case, with a quality of the fit
however, which is inferior to that shown on Fig.~\ref{fig:fitting}.

\section{Summary\label{sec:Summary}}

We have studied the specific heat and uniform susceptibility of mixed-spin
chain systems using a combination of high-temperature series expansion,
exact diagonalization, and quantum Monte Carlo techniques.
In particular, we have contrasted the cases of FM and AFM intra-chain exchange.
A symbolic high-temperature series has been derived for general values of the spin quantum numbers,
and including single ion anisotropies.
Using this series expansion,
we were able to extract the microscopic model parameters for
the quasi-one-dimensional FM mixed-spin chain compound MnNi(NO$_{2}$)$_{4}$(en)$_{2}$.
Comparing our results to the analytically solvable limit $S\rightarrow \infty $,
we found that not only the AFM but also the FM case
displays a double-peak-like structure in the specific heat which is due to the presence of
both `optical' and `acoustic' excitations. In fact, we find
that for FM intra-chain coupling this structure
is more pronounced than for AFM exchange.
The low-temperature specific heat of MnNi(NO$_{2}$)$_{4}$(en)$_{2}$
shows a weak shoulder around 1.5K if magnetic order is suppressed
by a finite magnetic field. It is thus suggestive to associate this shoulder
with the presence of `acoustic' excitations in this compound.
However, a quantitatively accurate description of the
relevant temperature range for the FM mixed-spin
chain with $s=1$ and $S=5/2$ in a magnetic field
requires further numerical studies
and is left for future investigations.
Finally, comparing the AFM with the FM case in the
extreme quantum limit, i.e. $s=1/2$, $S=1$, we found finite-size
effects  to be more pronounced in the FM than in the AFM
case. This suggests long-range collective excitations to be
more relevant for the low-temperature thermodynamics of FM mixed-spin chains.

\acknowledgments
The authors are grateful to R.~Feyerherm and S.~S\"{u}llow for
fruitful discussions about the experimental situation on MnNi(NO$_{2}$)$_{4}$(en)$_{2}$
and the provision of data. This work was supported in part by the
Deutsche Forschungsgemeinschaft through grant SU 229/6-1. S.W. acknowledges support from the Swiss National Science
Foundation. Parts of the numerical calculations were performed on the Asgard cluster at the ETH Z\"{u}rich
and on the {\tt cfgauss} at the computing center of the TU Braunschweig.

\appendix

{\small
\section{Series Data \label{sec:seriesdata}}

Using $X\equiv S(S+1)$ and $x\equiv s(s+1)$, the specific heat series
without the single-site anisotropy is given by \begin{equation}
C=C_{\textrm{D}}+C_{\textrm{ND}}(x,X)+C_{\textrm{ND}}(X,x)+O\left[(\beta J)^{12}\right],\end{equation}
 \begin{widetext}
\begin{eqnarray}
C_{\textrm{D}} &=&
\frac{2\,x\,X\,(\beta J)^{2}}{3} - \frac{x\,X\,(\beta J)^{3}}{3}
+ \left( \frac{2\,x\,X}{15} - \frac{2\,x^{2}\,X^{2}}{15} \right) \,(\beta J)^{4}   +
  \left( \frac{- x\,X }{18} + \frac{2\,x^{2}\,X^{2}}{27} \right) \,(\beta J)^{5}  \nonumber\\&& +
  \left( \frac{8\,x\,X}{315} + \frac{643\,x^{2}\,X^{2}}{7560} + \frac{4\,x^{3}\,X^{3}}{189} \right) \,(\beta J)^{6} +
  \left( \frac{-23\,x\,X}{1800} - \frac{4043\,x^{2}\,X^{2}}{32400} - \frac{2\,x^{3}\,X^{3}}{135}
\right) \,(\beta J)^{7} \nonumber\\&& +
  \left( \frac{19\,x\,X}{2700} + \frac{1162\,x^{2}\,X^{2}}{10125}
 - \frac{622\,x^{3}\,X^{3}}{18225} - \frac{2\,x^{4}\,X^{4}}{675} \right) \,
   (\beta J)^{8}  \nonumber\\&& + \left( \frac{-2231\,x\,X}{529200} - \frac{1489333\,x^{2}\,X^{2}}{15876000}
+ \frac{115043\,x^{3}\,X^{3}}{2976750} +
     \frac{4\,x^{4}\,X^{4}}{1575} \right) \,(\beta J)^{9}   \nonumber\\&& + \left( \frac{15901\,x\,X}{5821200}
 + \frac{15768563\,x^{2}\,X^{2}}{209563200} -
     \frac{5996723\,x^{3}\,X^{3}}{943034400} + \frac{36427\,x^{4}\,X^{4}}{3742200}
 + \frac{4\,x^{5}\,X^{5}}{10395} \right) \,(\beta J)^{10}  \nonumber\\&& +
  \left( \frac{-72557\,x\,X}{38102400} - \frac{822853\,x^{2}\,X^{2}}{13395375}
 - \frac{111661717\,x^{3}\,X^{3}}{3857868000} -
     \frac{1448899\,x^{4}\,X^{4}}{128595600} - \frac{2\,x^{5}\,X^{5}}{5103} \right) \,(\beta J)^{11},
\end{eqnarray}
\begin{eqnarray}
C_{\textrm{ND}}(x,X) &=& \frac{-8\,x\,X^{2}\,(\beta J)^{4}}{45} + \frac{4\,x\,X^{2}\,(\beta J)^{5}}{27} +
  \left( \frac{-179\,x\,X^{2}}{1890} + \frac{2\,x\,X^{3}}{63} + \frac{10\,x^{2}\,X^{3}}{189} \right) \,(\beta J)^{6}
 \nonumber\\&& +
  \left( \frac{157\,x\,X^{2}}{2700} - \frac{x\,X^{3}}{25} - \frac{82\,x^{2}\,X^{3}}{2025} \right) \,(\beta J)^{7}
 \nonumber\\&& +
  \left( \frac{-124\,x\,X^{2}}{3375} + \frac{124\,x\,X^{3}}{3375} - \frac{6091\,x^{2}\,X^{3}}{364500} - \frac{16\,x\,X^{4}}{3375} -
     \frac{1064\,x^{2}\,X^{4}}{91125} - \frac{8\,x^{3}\,X^{4}}{675} \right) \,(\beta J)^{8} \nonumber\\&& +
  \left( \frac{3229\,x\,X^{2}}{132300} - \frac{1019\,x\,X^{3}}{33075} + \frac{23281\,x^{2}\,X^{3}}{441000} + \frac{8\,x\,X^{4}}{945} +
     \frac{128\,x^{2}\,X^{4}}{10125} + \frac{244\,x^{3}\,X^{4}}{23625} \right) \,(\beta J)^{9} \nonumber\\&& +
  \left( \frac{-17137\,x\,X^{2}}{997920} + \frac{24727\,x\,X^{3}}{970200} - \frac{86018257\,x^{2}\,X^{3}}{1257379200}
   -
     \frac{11489\,x\,X^{4}}{1091475} + \frac{6287\,x^{2}\,X^{4}}{11642400}\right.\nonumber\\&&\left. + \frac{25411\,x^{3}\,X^{4}}{2619540} + \frac{4\,x\,X^{5}}{6237} +
     \frac{316\,x^{2}\,X^{5}}{155925} + \frac{218\,x^{3}\,X^{5}}{66825} + \frac{2\,x^{4}\,X^{5}}{891} \right) \,(\beta J)^{10} \nonumber\\&& +
  \left( \frac{732629\,x\,X^{2}}{57153600} - \frac{757\,x\,X^{3}}{35280} + \frac{11884871\,x^{2}\,X^{3}}{163296000} +
     \frac{10279\,x\,X^{4}}{893025} - \frac{5676361\,x^{2}\,X^{4}}{367416000} \right.\nonumber\\&&\left. - \frac{3918727\,x^{3}\,X^{4}}{214326000} -
     \frac{1382\,x\,X^{5}}{893025} - \frac{2672\,x^{2}\,X^{5}}{893025} - \frac{30181\,x^{3}\,X^{5}}{8037225} - \frac{32\,x^{4}\,X^{5}}{14175}
\right) \,(\beta J)^{11}.
\end{eqnarray}
\end{widetext}

We have computed the series of the susceptibility with the single-site anisotropy
up to up to $O(\beta^7)$, which is too lengthy to be fully listed in this paper.
Hence, we show here only the first four terms of the series,
and the rest will
be provided on request. Since the non-commutativity, Eq.~(\ref{notcommute}),
is neglected in Ref.~\onlinecite{Kahn92}, the $S\rightarrow \infty $
limit of the series below with $D=0$ and $g\neq G$
is different from the function given in
Ref.~\onlinecite{Kahn92}. \begin{widetext}
\begin{eqnarray}
\chi_{zz} &=&
\beta\,\left( \frac{{g}^{2}\,x}{3} + \frac{{G}^{2}\,X}{3} \right)  +
  {\beta}^{2}\,\left\{ \frac{4\,g\,G\,J\,x\,X}{9} -
     D\,{G}^{2}\,\left[ \frac{-X}{15} + \frac{4\,X^{2}}{45} \right]  \right\}  +
  {\beta}^{3}\,\Bigg\{ {g}^{2}\,J^{2}\,\left[ \frac{-\left( x\,X \right) }{27} + \frac{2\,x^{2}\,X}{27} \right]
 \nonumber\\&&
+
     g\,G\,\left[ \frac{-\left( J^{2}\,x\,X \right) }{27} -
        D\,J\,\left( \frac{-4\,x\,X}{45} + \frac{16\,x\,X^{2}}{135} \right)  \right]  +
     {G}^{2}\,\left[ J^{2}\,\left( \frac{-\left( x\,X \right) }{27} + \frac{2\,x\,X^{2}}{27} \right)  +
        {D}^{2}\,\left( \frac{X}{42} - \frac{4\,X^{2}}{105} + \frac{8\,X^{3}}{945} \right)  \right]  \Bigg\}
 \nonumber\\&&
 +
  {\beta}^{4}\,\Bigg\{ {g}^{2}\,\left[ J^{3}\,\left( \frac{x\,X}{108} - \frac{x^{2}\,X}{81} \right)  -
        D\,J^{2}\,\left( \frac{2\,x\,X}{675} - \frac{16\,x^{2}\,X}{675} - \frac{8\,x\,X^{2}}{2025} +
           \frac{64\,x^{2}\,X^{2}}{2025} \right)  \right]
 +
     g\,G\,\left[ D\,J^{2}\,\left( \frac{4\,x\,X^{2}}{405}  - \frac{x\,X}{135}  \right)
 \right.\nonumber\\&&\left.
 +
        J^{3}\,\left( \frac{x\,X}{90} - \frac{16\,x^{2}\,X}{405} - \frac{16\,x\,X^{2}}{405} + \frac{8\,x^{2}\,X^{2}}{405} \right) +
        {D}^{2}\,J\,\left( \frac{2\,x\,X}{63} - \frac{16\,x\,X^{2}}{315} + \frac{32\,x\,X^{3}}{2835} \right)  \right]
 \nonumber\\&&
 +
     {G}^{2}\,\left[ J^{3}\,\left( \frac{x\,X}{108} - \frac{x\,X^{2}}{81} \right)  -
        D\,J^{2}\,\left( \frac{x\,X}{54} - \frac{22\,x\,X^{2}}{405} + \frac{16\,x\,X^{3}}{405} \right)  -
        {D}^{3}\,\left( \frac{-X}{90} + \frac{97\,X^{2}}{4725} - \frac{32\,X^{3}}{4725} - \frac{16\,X^{4}}{14175} \right)
        \right]  \Bigg\}
\nonumber\\&&
 + \ldots,
\end{eqnarray}
\begin{eqnarray}
\chi_{xx}  &=&
\beta\,\left( \frac{{g}^{2}\,x}{3} + \frac{{G}^{2}\,X}{3} \right)  +
  {\beta}^{2}\,\left\{ \frac{4\,g\,G\,J\,x\,X}{9} -
     D\,{G}^{2}\,\left( \frac{X}{30} - \frac{2\,X^{2}}{45} \right)  \right\}  +
  {\beta}^{3}\,\Bigg\{ {g}^{2}\,J^{2}\,\left[ \frac{-\left( x\,X \right) }{27} + \frac{2\,x^{2}\,X}{27} \right]
\nonumber\\&&
 +
     g\,G\,\left[ \frac{-\left( J^{2}\,x\,X \right) }{27} -
        D\,J\,\left( \frac{2\,x\,X}{45} - \frac{8\,x\,X^{2}}{135} \right)  \right]  +
     {G}^{2}\,\left[ J^{2}\,\left( \frac{-\left( x\,X \right) }{27} + \frac{2\,x\,X^{2}}{27} \right)  +
        {D}^{2}\,\left( \frac{X}{210} - \frac{X^{2}}{315} - \frac{4\,X^{3}}{945} \right)  \right]  \Bigg\}
\nonumber\\&&
 +
  {\beta}^{4}\,\Bigg\{ {g}^{2}\,\left[ J^{3}\,\left( \frac{x\,X}{108} - \frac{x^{2}\,X}{81} \right)  -
        D\,J^{2}\,\left( \frac{-\left( x\,X \right) }{675} + \frac{8\,x^{2}\,X}{675} + \frac{4\,x\,X^{2}}{2025} -
           \frac{32\,x^{2}\,X^{2}}{2025} \right)  \right]
\nonumber\\&&
+
     g\,G\,\left[ D\,J^{2}\,
         \left( \frac{x\,X}{270} - \frac{2\,x\,X^{2}}{405} \right)  +
        J^{3}\,\left( \frac{x\,X}{90} - \frac{16\,x^{2}\,X}{405} - \frac{16\,x\,X^{2}}{405} + \frac{8\,x^{2}\,X^{2}}{405} \right)
\right.\nonumber\\&&\left.
 +
        {D}^{2}\,J\,\left( \frac{2\,x\,X}{315} - \frac{4\,x\,X^{2}}{945} - \frac{16\,x\,X^{3}}{2835} \right)  \right]  +
     {G}^{2}\,\left[ J^{3}\,\left( \frac{x\,X}{108} - \frac{x\,X^{2}}{81} \right)  -
        D\,J^{2}\,\left( \frac{-\left( x\,X \right) }{108} + \frac{11\,x\,X^{2}}{405} - \frac{8\,x\,X^{3}}{405} \right)
\right.\nonumber\\&&\left.
-
        {D}^{3}\,\left( \frac{-X}{2520} - \frac{X^{2}}{1350} + \frac{2\,X^{3}}{1575} + \frac{8\,X^{4}}{14175} \right)  \right]
     \Bigg\} + \ldots.
\end{eqnarray}

\end{widetext}
}


\begin{thebibliography}{10}
\bibitem{finitefield}In a finite field, the correction term is replaced by $O[(\beta J)^{\ell }]$,
and the other terms are the same.
\bibitem{misprint}The series coefficients listed in Fig.~2 in Ref.~\onlinecite{Wojtowicz67}
are not consistent with that of
the uniform system\cite{Baker64,Rushbrooke58}, i.e. the $s=S$ case.
Most likely, ``$\pm3200$''
in the table should be replaced by ``$\pm4200$''.
\end{thebibliography}
\end{document}